\documentclass[pra,twocolumn,tightenlines,showpacs,nofootinbib
]{revtex4}
\usepackage{bm,dcolumn,amsmath,graphicx}

\newcommand{\Eref}[1]{Eq.~(\ref{#1})}
\newcommand{\tref}[1]{Table~\ref{#1}}

\begin{document}

\title{Transition frequency shifts with fine structure
constant variation for {Fe~II}: Breit and core-valence correlation
corrections}

\author{S. G. Porsev$^{1}$}
\author{K. V. Koshelev$^{1,2}$}
\author{I. I. Tupitsyn$^3$}
\author{M. G. Kozlov$^1$}
\author{D. Reimers$^2$}
\author{S.~A. Levshakov$^4$}
\affiliation{$^1$ Petersburg Nuclear Physics Institute, Gatchina,
188300, Russia}
\affiliation{$^2$~Hamburger Sternwarte, Universit\"{a}t Hamburg, Hamburg, Germany}
\affiliation{$^3$~St.Petersburg State University, Petrodvorets, Russia}
\affiliation{$^4$~Ioffe Physico-Technical Institute, St. Petersburg, Russia}

\date{ \today }
\pacs{31.30.Jv, 06.20.Jr}

\begin{abstract}
Transition frequencies of Fe~\textsc{ii} ion are known to be very
sensitive to variation of the fine structure constant $\alpha$. The
resonance absorption lines of Fe~\textsc{ii} from objects at
cosmological distances are used in a search for the possible
variation of $\alpha$ in cause of cosmic time. In this paper we
calculated the dependence of the transition frequencies on
$\alpha^2$ ($q$-factors) for Fe~\textsc{ii} ion. We found
corrections to these coefficients from valence-valence and
core-valence correlations and from the Breit interaction. Both the
core-valence correlation and Breit corrections to the $q$-factors
appeared to be larger than had been anticipated previously.
Nevertheless our calculation confirms
that the Fe~\textsc{ii} absorption lines seen in quasar spectra
have large $q$-factors of both signs
and thus the ion Fe~\textsc{ii} alone can be used in the search for the
$\alpha$-variation at different cosmological epochs.
\end{abstract}

\maketitle

\section{Introduction}\label{sec_intro}

Recently discovered acceleration of the universe(for a review, see
\cite{Cop06}) is usually regarded as evidence for the existence of
dark energy. Cosmological evolution of the dark energy may cause
variations of fundamental constants, such as the fine-structure
constant $\alpha$ and proton to electron mass ratio $\mu=m_p/m_e$
(see, e.g., \cite{Fla07} and references therein). Different models
predict different behavior of the coupling constants and it is
extremely important to measure $\alpha$ at different cosmological
epochs\footnote{Each cosmological epoch is characterized by the
redshift parameter $z$, which is defined in the spectral
observations as $z = (\lambda_{\rm obs} - \lambda_{\rm
lab})/\lambda_{\rm lab}$, with $\lambda_{\rm obs}, \lambda_{\rm
lab}$ being correspondingly the observational and laboratory
wavelengths of an atomic transition. Currently we observe objects
with $z$ ranging from 0 (local universe) up to 7.}.

Because of the relativistic effects in atoms their transition
frequencies depend on $\alpha Z$, where $Z$ is atomic number.
Therefore, one can study space-time variation of $\alpha$ by
comparing atomic frequencies for distant objects in the universe
with their laboratory values. To do this one needs to calculate so
called $q$-factors for the atomic transitions which are defined as
follows:
 \begin{align}\label{qfactor1}
 \omega = \omega_\mathrm{lab} + q x, \quad x \equiv
 \left({\alpha}/{\alpha_\mathrm{lab}}\right)^2 - 1\,.
 \end{align}
Generally $q$-factors may have different signs and scale with atomic
number as $Z^2$ (see \cite{DFW99a}).
That allows to look for $\alpha$-variation by
comparing transition frequencies of heavy
and light ions \cite{MWF03,SCP04}.
However, under astrophysical conditions the relative fractions
of different ions depend on the gas density fluctuations within
the absorbing cloud and thus the measured positions of the absorption
lines are affected by irregular Doppler velocity shifts of the bulk
motions leading to an additional noise in the $\Delta\alpha/\alpha$
measurements \cite{L07, M07}.


Therefore, one can try to find an atom, where $q$-factors for
different transitions have different signs and large absolute
values. In Ref.~\cite{DFK02} it was shown, that the line 1608~\AA\
in Fe~\textsc{ii} has large negative $q$-factor, while other UV
resonance lines have large positive $q$-factors. Additional
advantage of Fe~\textsc{ii} is relatively small values of isotope
shifts \cite{KKBD04}. This fact was used in \cite{L07,L06,L05,Q04}
to suggest the Single Ion Differential $\alpha$ Measurement (SIDAM)
procedure which is less sensitive to uncertainties inherent to the
method applied in \cite{MWF03,SCP04} and allows us to measure
$\Delta\alpha/\alpha$ at a single redshift $z$ with a sufficiently
high accuracy ($\sigma_{\Delta\alpha/\alpha} \sim 10^{-6}-10^{-5}$).

Here we report new calculations of the $q$-factors for
Fe~\textsc{ii} with more accurate account for relativistic effects
and electronic correlations. In particular, we studied corrections
to the $q$-factors from the Breit interaction and the core-valence
correlations. We found that both corrections are noticeable but do
not change previous conclusion~\cite{DFK02} that Fe~\textsc{ii} has
large $q$-factors of both signs.

\section{Method of calculation}\label{sec_method}

In order to calculate $q$-factors one can solve atomic relativistic
eigenvalue problem for different values of $\alpha$. For example, if
we calculate atomic frequency $\omega_\pm$ for two values $x=\pm
1/8$ of the parameter $x$ in \Eref{qfactor1}, the corresponding
$q$-factor is given by
 \begin{align}\label{qfactor2}
  q=4(\omega_+-\omega_-).
 \end{align}
The choice of $x=\pm 1/8$ is usually sufficiently small to neglect
nonlinear corrections and sufficiently large to make calculations
numerically stable. The exceptions may occur for the strongly
interacting levels. Such cases require special consideration.

Fe~\textsc{ii} ion has ground state configuration [Ar]$3d^6 4s$ with
7 electrons in open shells, hence it has rather dense and
complicated spectrum, particularly in the astrophysically important
region above $60000~\mathrm{cm}^{-1}$ from the ground state.
Calculations for Fe~\textsc{ii} are very difficult because of the
large number of valence electrons and high density of the spectrum
\cite{DFW99a,DFK02,Bec05,Bec07}.

We use Dirac-Coulomb and Dirac-Coulomb-Breit ``no pair''
Hamiltonians. At first stage we use the frozen-core approximation
and solve valence eigenvalue problem by means of configuration
interaction (CI) method. Using \Eref{qfactor2} it is important that
CI space is the same for two values of $x$. When four-component
finite basis sets are used to solve many-electron Dirac equations,
the basic orbitals depend on $\alpha$. The incompleteness of the CI
space can introduce some error for $q$-factors. To check if this
error is small we make several CI calculations with different basis
sets.

At first we solve Dirac-Fock equations with the code \cite{BDT77} to
find core orbitals $1s,\dots,3p_{3/2}$ and valence orbitals
$3d_{3/2},3d_{5/2},4s,4p_{1/2},4p_{3/2}$. Then we add virtual
orbitals, which are constructed using the method described in
\cite{Bog91,KPF96}. In this method an upper component of virtual
orbitals is formed from the previous orbital of the same symmetry by
multiplication by some monotone function of radial variable $r$. The
lower component is then formed using kinetic balance condition. (An
alternative method to form basis sets was suggested in \cite{TL02}).
Our basis sets include $s,p,d,$ and $f$ orbitals with principle
quantum number $n\le N$ and are designated as $[Nspdf]$.
Configuration space is formed by single and double (SD) excitations
from the configurations $3d^6 4s$, $3d^6 4p$, and $3d^5 4s4p$.
Respective dimensions of the CI space are significantly larger than
in previous calculations \cite{DFW99a,DFK02}.

In order to study Breit correction to the $q$-factors we do
calculations for Coulomb, Coulomb-Gaunt and Coulomb-Breit
approximations, where Gaunt approximation accounts for the magnetic
interaction and Breit interaction also includes retardation term. We
have found that due to the large number of valence electrons and
relative smallness of the nuclear charge $Z$ it is necessary to
include Breit correction to the valence-valence interactions as well
as to the core-valence and core-core ones.

In our calculation the Ar-like core is rather small and rigid, with
binding energies $|\varepsilon| > 3$~a.u. For this reason one can
expect that corrections from core excitations should be small. The
same conclusion follows from the relatively good agreement between
theoretical spectra in the frozen core approximation and the
experiment \cite{NIST}. On the other hand, correlations between core
electrons and valence $3d$-electrons are enhanced. For this reason
the core-valence correlations require thorough investigation.

Usually one can use the many-body perturbation theory (MBPT) to
account for core-valence correlations. Unfortunately, it is
difficult to implement MBPT for atoms with open $d$-shells because
valence $d$-shell electrons belong to the previous atomic shell.
That leads to much stronger interaction between valence
$d$-electrons and uppermost core electrons \cite{DJ98,KP99,SSK07}.
Because of that it is impractical to use the $V^{N_\mathrm{core}}$
approximation, where MBPT is much simpler than in the case of
$V^{N}$ potential with $N\ne N_\mathrm{core}$. Generally MBPT
includes a large class of the so-called subtraction diagrams, which
cancel out for $V^{N_\mathrm{core}}$ approximation. That not only
increases the complexity of calculations but also makes them less
stable because of the large cancelations between ``normal'' and
subtraction diagrams \cite{Koz03}. Due to these problems the CI+MBPT
method \cite{DFK96b}, which proved to be rather effective for
treating atoms with several $s$- and $p$-electrons in the open
shells \cite{PKRD01,KPJ01,Dzu05a,Dzu05b,DF07a}, is much less
effective here. Because of that we combine CI+MBPT method with
conventional CI calculation for 15 electrons including 8 outermost
core electrons.

\section{Results}\label{results}

\subsection{Valence CI}\label{val_CI}

The results of the 7-electron CI calculation of the spectrum of
Fe~\textsc{ii} are listed in \tref{tab_w_g_q}. The CI space for each
basis set corresponds to SD excitations from configurations $3d^6
4s$, $3d^6 4p$, and $3d^5 4s4p$. It is convenient to divide all
levels presented in~\tref{tab_w_g_q} into three groups. The first
group includes five levels of the even-parity $^6D_J$ manifold. To
the second group we relate the low-lying odd-parity states of the
configuration $3d^6 4p$, namely $^6D^o_{9/2,7/2}$,
$^6F^o_{11/2,9/2}$, and $^6P^o_{7/2}$. Finally, in the third group
we include two levels of the configuration $3d^5 4s4p$
($^8P^o_{7/2}$ and $^6P^o_{7/2}$) and the closely located states of
the configuration $3d^6 4p$: $^4G^o_{7/2}$, $^4H^o_{7/2}$, and
$^4F^o_{7/2}$.

\begin{table*}[bt]
\caption{Seven-electron CI calculations of transition frequencies
$\omega$ (in cm$^{-1}$), $g$-factors, and $q$-factors (in cm$^{-1}$)
for different basis sets. Dirac-Coulomb Hamiltonian in the
frozen-core approximation is used. Transition frequencies and
$q$-factors are calculated in respect to the ground state.}

\label{tab_w_g_q}

\begin{ruledtabular}

\begin{tabular}{llrlrlrrlrrlrrlr}
&&\multicolumn{2}{c}{Exper.~\cite{NIST}}
&\multicolumn{3}{c}{$[5spdf]$} &\multicolumn{3}{c}{$[6spdf]$}
&\multicolumn{3}{c}{$[7spdf]$}
&\multicolumn{3}{c}{Theory~\cite{DFK02}\footnotemark[1]}

\\
Conf.& Level &\multicolumn{1}{c}{$\omega$}&\multicolumn{1}{c}{$g$}
&\multicolumn{1}{c}{$\omega$}&\multicolumn{1}{c}{$g$}&\multicolumn{1}{c}{$q$}
&\multicolumn{1}{c}{$\omega$}&\multicolumn{1}{c}{$g$}&\multicolumn{1}{c}{$q$}
&\multicolumn{1}{c}{$\omega$}&\multicolumn{1}{c}{$g$}&\multicolumn{1}{c}{$q$}
&\multicolumn{1}{c}{$\omega$}&\multicolumn{1}{c}{$g$}&\multicolumn{1}{c}{$q$}
\\[1mm]
\hline
$3d^6 4s$   & $^6D_{9/2}$   &     0 & 1.58  &     0 & 1.555 &$    0 $&     0 & 1.556 &$    0 $&     0 & 1.555 &$    0 $&    0  &       &$    0 $\\
$3d^6 4s$   & $^6D_{7/2}$   &   385 & 1.58  &   382 & 1.587 &$  384 $&   376 & 1.586 &$  378 $&   375 & 1.587 &$  377 $&       &       &$      $\\
$3d^6 4s$   & $^6D_{5/2}$   &   668 & 1.665 &   666 & 1.657 &$  658 $&   656 & 1.657 &$  647 $&   653 & 1.656 &$  644 $&       &       &$      $\\
$3d^6 4s$   & $^6D_{3/2}$   &   863 & 1.862 &   864 & 1.866 &$  843 $&   849 & 1.866 &$  829 $&   846 & 1.866 &$  825 $&       &       &$      $\\
$3d^6 4s$   & $^6D_{1/2}$   &   977 & 3.31  &   980 & 3.332 &$  950 $&   963 & 3.333 &$  934 $&   960 & 3.332 &$  930 $&       &       &$      $\\[1mm]
$3d^6 4p$   & $^6D^o_{9/2}$ & 38459 & 1.542 & 37218 & 1.554 &$ 1334 $& 37366 & 1.554 &$ 1340 $& 37370 & 1.554 &$ 1334 $& 38352 &       &$ 1361 $\\
$3d^6 4p$   & $^6D^o_{7/2}$ & 38660 & 1.584 & 37424 & 1.586 &$ 1491 $& 37567 & 1.586 &$ 1492 $& 37570 & 1.586 &$ 1488 $& 38554 & 1.586 &$ 1516 $\\
$3d^6 4p$   & $^6F^o_{11/2}$& 41968 &       & 40933 & 1.454 &$ 1478 $& 41094 & 1.454 &$ 1485 $& 41095 & 1.454 &$ 1479 $& 41864 &       &$ 1502 $\\
$3d^6 4p$   & $^6F^o_{9/2}$ & 42115 & 1.43  & 41088 & 1.433 &$ 1608 $& 41245 & 1.433 &$ 1612 $& 41245 & 1.433 &$ 1607 $& 42012 &       &$ 1623 $\\
$3d^6 4p$   & $^6P^o_{7/2}$ & 42658 & 1.702 & 41562 & 1.708 &$ 1140 $& 41746 & 1.709 &$ 1136 $& 41758 & 1.709 &$ 1134 $& 42715 & 1.709 &$ 1251 $\\[1mm]
$3d^5 4s4p$ & $^8P^o_{7/2}$ & 52583 &       & 48817 & 1.936 &$-2058 $& 48919 & 1.936 &$-2090 $& 49113 & 1.936 &$-2103 $& 54914 & 1.936 &$-2085 $\\
$3d^6 4p$   & $^4G^o_{7/2}$ & 60957 & 0.969 & 62604 & 0.970 &$ 1521 $& 62789 & 0.962 &$ 1468 $& 62773 & 0.973 &$ 1485 $& 63624 & 0.978 &$ 1640 $\\
$3d^6 4p$   & $^4H^o_{7/2}$ & 61157 & 0.720 & 62578 & 0.715 &$ 1232 $& 62861 & 0.726 &$ 1228 $& 62900 & 0.718 &$ 1171 $& 63498 & 0.703 &$ 1272 $\\
$3d^6 4p$   & $^4F^o_{7/2}$ & 62066 & 1.198 & 64299 & 1.227 &$ 1166 $& 64307 & 1.213 &$ 1270 $& 64023 & 1.208 &$ 1240 $& 65528 & 1.252 &$ 1062 $\\
$3d^5 4s4p$ & $^6P^o_{7/2}$ & 62172 & 1.68  & 59011 & 1.714 &$-1457 $& 59045 & 1.714 &$-1491 $& 59242 & 1.714 &$-1506 $& 65750 & 1.713 &$-1519 $\\
\end{tabular}
\end{ruledtabular}
\footnotemark[1]{These $q$-factors do not include semi-empirical
corrections from fitting $g$-factors. The respective corrections are
discussed in Sec.~\ref{level6P7}.}
\end{table*}

Properties of the even-parity levels from the first group (fine
structure splitting, $g$- and $q$-factors) are rather insensitive to
configuration interaction. All these quantities change very weakly
with increasing the basis set and the size of the configuration
space. An explanation is that the weight of the leading
configuration $3d^6 4s$ for these states is about 95-97\%. Small
admixture of other configurations does not influence their
properties.

Behavior of the odd levels from the second group is rather similar.
It is seen from~\tref{tab_w_g_q} that $g$-factors of these levels
remain practically the same while transition frequencies and
$q$-factors change by less than 4\% when the basis set is increased
from $[5spdf]$ to $[7spdf]$. The weight of the leading configuration
$3d^6 4p$ for the $^6P^o_{7/2}$ state is about 90\% and greater than
95\% for other states. The levels of different multiplets with the
same total angular momentum $J$ are well separated and interact with
each other rather weakly.

Comparing calculated transition frequencies for this group of levels
with experimental data we see very good agreement. Largest
discrepancy does not exceed 3\%. There is also a reasonable
agreement between the $q$-factors obtained in this work and in
Ref.~\cite{DFK02}. Except for the state $3d^6 4p\, {}^6\!P^o_{7/2}$
the differences between $q$-factors are at the level of 1-2\%. For
the state $3d^6 4p\,{}^6\!P^o_{7/2}$ the difference is about 5\%.
The explanation for this difference is probably the following.
Calculation in the paper~\cite{DFK02} was done on the basis set
$[6spdf]$ and the CI space included SD excitations from the
configuration $3d^6 4p$. In order to accelerate calculations this
space was then truncated by excluding non-relativistic
configurations, whose weight in the levels of interest was smaller
than 0.5\%. Here we do not truncate configuration space and also
include the SD excitations from the configuration $3d^5 4s4p$, which
corresponds to triple excitations from the configuration $3d^6 4p$.
Admixture of the configuration $3d^5 4s4p$ to the state $3d^6 4p\,
{}^6\!P^o_{7/2}$ is at the level of 5\%. For this reason the SD
excitations from the configuration $3d^5 4s4p$ appear to be more
important for this level.


\tref{tab_w_g_q} shows that transition frequencies for the third
group of levels also rather weakly depend on the size of the CI
space. Comparing the results obtained for the basis sets $[5spdf]$
and $[7spdf]$ we see that a most significant change of $\omega$
($\sim$ 320 cm$^{-1}$) is for the $^4H^o_{7/2}$ state. Even here it
constitutes only 0.5\% of the transition frequency. On the other
hand, because of the high density of levels in this part of the
spectrum, such changes are comparable to the spacings between
different levels with the same $J$. The Lande-factors $g$ for
several levels with $J=7/2$ of the configuration $3d^6 4p$ are
changing rather strongly from one calculation to another. That
signals that these levels interact with each other. However, this
interaction does not affect the $q$-factors too much since all these
levels have similar dependence on $\alpha$. The Lande-factors of the
levels of the configuration $3d^5 4s4p$, in contrast, are very
stable. For the most interesting level $^6P^o_{7/2}$ theoretical
$g$-factor is somewhat larger than in the experiment. This will be
discussed in detail in Sec.~\ref{level6P7}.

We can sum up, that on the stage of the valence CI calculation there
is agreement on the level of few percent between theoretical and
experimental frequencies. However, the theory does not reproduce the
intervals between levels in the higher part of the spectrum, where
the density of states is very high. We see that theoretical
frequencies do not change much with the size of the valence space.
We can therefore conclude that we are rather close to saturation of
the valence CI space and the remaining difference between theory and
experiment is mostly caused by neglect of the core-valence
correlations. We will discuss this part of the problem in
Sec.~\ref{core}.

\subsection{Breit corrections}\label{Breit}

The results presented in \tref{tab_w_g_q} correspond to the
Dirac-Coulomb approximation. As long as we are mostly interested in
$q$-factors, which depend on the relativistic corrections to
electronic Hamiltonian, the role of the Breit interaction is
enhanced. In order to estimate corresponding corrections we made
Dirac-Coulomb and Dirac-Coulomb-Breit calculations of the
$q$-factors in the one-configurational approximation. The results of
these calculations are given in \tref{tab_breit}.

\begin{table}[htb]
\caption{One-configurational calculation of $q$-factors in
Dirac-Coulomb, Dirac-Coulomb-Gaunt, and Dirac-Coulomb-Breit
approximations (in cm$^{-1}$); $\delta q_{gn}$ and $\delta q_{br}$
are respective Gaunt and Breit corrections to Dirac-Coulomb
$q$-factors $q_{DC}$. The latter are calculated for $x=\pm 1/8$
using \Eref{qfactor2} and also for $x=\pm 0.01$.}

\label{tab_breit}

\begin{ruledtabular}

\begin{tabular}{llrrrr}
&&\multicolumn{2}{c}{$q_{DC}$}\\
Conf.& Level &\multicolumn{1}{c}{$|x|=1/8$}&\multicolumn{1}{c}{$|x|=0.01$}
&\multicolumn{1}{c}{$\delta q_{gn}$}
&\multicolumn{1}{c}{$\delta q_{br}$}\\
\hline
$3d^6 4s$   & $^6D_{9/2}$   &$    0.0 $&$    0.0 $&$   0.0 $&$   0.0 $\\       
$3d^6 4s$   & $^6D_{7/2}$   &$  377.0 $&$  377.1 $&$ -23.2 $&$ -23.3 $\\       
$3d^6 4s$   & $^6D_{5/2}$   &$  648.5 $&$  648.6 $&$ -39.2 $&$ -39.3 $\\       
$3d^6 4s$   & $^6D_{3/2}$   &$  833.1 $&$  833.2 $&$ -49.6 $&$ -49.8 $\\       
$3d^6 4s$   & $^6D_{1/2}$   &$  940.7 $&$  940.8 $&$ -55.6 $&$ -55.7 $\\[1mm]  
$3d^6 4p$   & $^6D^o_{9/2}$ &$ 1296.3 $&$ 1296.4 $&$  22.1 $&$  18.7 $\\       
$3d^6 4p$   & $^6D^o_{7/2}$ &$ 1475.8 $&$ 1475.9 $&$  14.7 $&$  11.2 $\\       
$3d^6 4p$   & $^6F^o_{11/2}$&$ 1438.6 $&$ 1440.8 $&$  13.4 $&$  10.6 $\\
$3d^6 4p$   & $^6F^o_{9/2}$ &$ 1614.6 $&$ 1616.9 $&$  -0.3 $&$  -3.1 $\\
$3d^6 4p$   & $^6P^o_{7/2}$ &$ 1331.8 $&$ 1333.9 $&$  15.7 $&$  13.2 $\\[1mm]  
$3d^5 4s4p$ & $^8P^o_{7/2}$ &$-2433.6 $&$-2433.8 $&$ 175.4 $&$ 156.0 $\\       
$3d^6 4p$   & $^4G^o_{7/2}$ &$ 1617.4 $&$ 1619.9 $&$ -55.2 $&$ -58.1 $\\
$3d^6 4p$   & $^4H^o_{7/2}$ &$ 1334.5 $&$ 1336.5 $&$  66.9 $&$  63.4 $\\
$3d^6 4p$   & $^4F^o_{7/2}$ &$ 1172.5 $&$ 1173.4 $&$  49.8 $&$  46.6 $\\
$3d^5 4s4p$ & $^6P^o_{7/2}$ &$-1824.7 $&$-1824.9 $&$ 148.1 $&$ 130.7 $\\       
\end{tabular}
\end{ruledtabular}
\end{table}

One can see that the Breit corrections to the $q$-factors are non
negligible, particularly for the levels of the configuration $3d^5
4s4p$. It is worth mentioning that, as usual, the magnetic part of
the Breit interaction (Gaunt interaction) is significantly larger
than the retardation part. On the other hand, the Breit corrections
to the valence-valence interactions are comparable to the
corrections to core-valence interactions. The reason for that is
relatively small size of the core and the large number of valence
$3d$-electrons. Because of that it is important to include Breit
interaction not only in the mean field potential, but also in the
residual two-electron interaction. On the other hand, the Breit
corrections to the $q$-factors are sufficiently small (i.e. less
than 10\%), so that they can be calculated within
one-configurational approximation as in \tref{tab_breit}. Below we
will add this correction to our final result in \tref{tab_final}.

In \tref{tab_breit} we also checked the size of the nonlinear
corrections from using \Eref{qfactor2} for evaluation of the
$q$-factors. For the Dirac-Coulomb approximation we calculated
$q$-factors using $x=\pm 1/8$ and $x=\pm 0.01$. One can see that the
differences between two calculations are on the order of a fraction
of a percent. We conclude that nonlinear corrections to
\Eref{qfactor2} can be neglected for all levels considered here.

\subsection{Core-valence correlations}\label{core}

As we mentioned in Sec.~\ref{sec_intro} the size of the core-valence
correlation corrections \textit{a priori} is not clear. On the one
hand the Ar-like core for Fe~\textsc{ii} is rather small and rigid.
On the other hand valence $3d$-electrons belong to the shell with
the same principal quantum number as the outermost core electrons
$3s,3p$. Thus, there is no spatial separation between core and
valence electrons.

There are two approaches that can be applied to account for
core-valence correlations. The first approach is to use the CI+MBPT
method \cite{DFK96b} and another one is to include two $3s$- and six
$3p$-electrons into the valence space. The former method allows
using long basis sets and accounting for correlations with all core
shells. The main problem with this method is instability of the MBPT
for the mean-field potential $V^N$, which includes large number of
valence electrons, i.e. for $N-N_\mathrm{core}\gg 1$. At present
this method is implemented only for the second order MBPT
corrections to valence Hamiltonian. That may be insufficient for
accurate treatment of correlations of valence $3d$-electrons with
$3s$- and $3p$-electrons from the core. It is known that second
order MBPT usually overestimates core-valence correlations.
Potentially more accurate all-order methods, like the one suggested
in Ref.~\cite{Koz04}, are currently used only for the systems with
two valence electrons. For example, coupled-cluster calculations of
the $q$-factors of a number of divalent atoms and ions was recently
reported in Ref.~\cite{BEIK06}.

The latter method is more straightforward, but the problem here is
the enormous size of the CI space even for rather small basis set.
Because of that it is impossible to saturate CI, which leads to
significant underestimation of the core-valence correlations. Due to
these problems neither method can be preferred and we used their
combination to estimate corrections caused by core-valence
correlations.

We started with solving 7-electron CI problem for the basis set
$[5s4pd]$. Then we solved 15-electron CI problem using the same
basis set and found correction $\delta q_\mathrm{cv,1}$, which
accounted for the correlations between valence electrons and $3s$-
and $3p$-electrons from the core. In both cases CI space included SD
excitations. The dimension of the CI matrix for the 15-electron
problem was on the order of $10^6$. That limited the length of the
basis set we could use in this approach. Such a short basis set was
obviously insufficient for accurate treatment of core-valence
correlations. To account for the incompleteness of the basis set we
performed two calculations of the $q$-factors using CI+MBPT method.
In one calculation all summations over virtual states in MBPT
diagrams ran over the same basis set as above and sums over core
states included only $3s$- and $3p$-shells.

Another calculation used a much longer basis set
$[21s\,16p\,21d\,19f\,14g]$ and all core shells. Valence space for
both calculations was the same as in the 7-electron CI above. The
difference between these two calculations gave us the second
correction $\delta q_\mathrm{cv,2}$, which accounted for
correlations with the inner core shells and the incompleteness of
the basis set used in 15-electron CI. The results of these
calculations are presented in \tref{tab_core}.

In the 15-electron CI calculation the odd-parity levels
$^4\!G_{7/2}^o$ and $^4\!H_{7/2}^o$ strongly interacted with each
other and their $g$-factors significantly differed from the
7-electron calculation and from the experiment. That hampered
comparison of the respective $q$-factors. We excluded these levels
from \tref{tab_core}.

In general CI+MBPT even for a short basis set gave much larger
corrections to $q$-factors and to transition frequencies than
15-electron CI. Comparison of the transition frequencies with the
experiment showed that CI+MBPT method strongly overestimated the
core-valence correlations. On the other hand, the 15-electron CI was
much closer to the experiment. Because of that we used CI+MBPT
method only to account for the incompleteness of the basis set. Our
final estimate of the correction from the core-valence correlation
is given by the sum $\delta q_\mathrm{cv}=\delta q_\mathrm{cv,1}
+\delta q_\mathrm{cv,2}$.

\begin{table}[htb]
\caption{Core-valence corrections to $q$-factors, $\delta
q_\mathrm{cv}=\delta q_\mathrm{cv,1} + \delta q_\mathrm{cv,2}$ (in
cm$^{-1}$). The term $\delta q_\mathrm{cv,1}$ is a difference of
15-electron and 7-electron CI on the basis set $[5s4pd]$. The second
term accounts for the incompleteness of this basis set. It is
obtained as a difference of 7-electron CI+MBPT on the basis sets
$[21s\,16p\,21d\,19f\,14g]$ and $[5s4pd]$.}

\label{tab_core}

\begin{ruledtabular}
\begin{tabular}{llrrr}
Conf.& Level
&\multicolumn{1}{c}{$\delta q_\mathrm{cv,1}$}
&\multicolumn{1}{c}{$\delta q_\mathrm{cv,2}$}
&\multicolumn{1}{c}{$\delta q_\mathrm{cv}$}
\\
\hline
$3d^6 4s$   & $^6D_{9/2}$   &$   0.0$&$   0.0 $&$   0.0 $\\         
$3d^6 4s$   & $^6D_{7/2}$   &$  10.5$&$  12.3 $&$  22.8 $\\         
$3d^6 4s$   & $^6D_{5/2}$   &$  11.4$&$  25.5 $&$  36.9 $\\         
$3d^6 4s$   & $^6D_{3/2}$   &$  10.5$&$  36.9 $&$  47.4 $\\         
$3d^6 4s$   & $^6D_{1/2}$   &$   8.8$&$  43.0 $&$  51.8 $\\[1mm]    
$3d^6 4p$   & $^6D^o_{9/2}$ &$  19.3$&$  39.5 $&$  58.8 $\\         
$3d^6 4p$   & $^6D^o_{7/2}$ &$  16.7$&$  19.3 $&$  36.0 $\\         
$3d^6 4p$   & $^6F^o_{11/2}$&$  17.6$&$  37.7 $&$  55.3 $\\         
$3d^6 4p$   & $^6F^o_{9/2}$ &$  19.3$&$  36.0 $&$  55.3 $\\         
$3d^6 4p$   & $^6P^o_{7/2}$ &$ 357.3$&$  36.9 $&$ 394.2 $\\[1mm]    
$3d^5 4s4p$ & $^8P^o_{7/2}$ &$  51.8$&$  99.2 $&$ 151.0 $\\         
$3d^6 4p$   & $^4F^o_{7/2}$ &$ 240.5$&$ 208.1 $&$ 448.6 $\\         
$3d^5 4s4p$ & $^6P^o_{7/2}$ &$  69.4$&$  98.3 $&$ 167.7 $\\         
\end{tabular}
\end{ruledtabular}
\end{table}

\subsection{Level ${}^6\!P^o_{7/2}$ at 62172 cm$^{-1}$}\label{level6P7}

The level $^6P^o_{7/2}$ at 62172 cm$^{-1}$ is the most important
level for the search of $\alpha$-variation. This one is the only
level of the configuration $3d^5 4s4p$, which is observed in
astrophysics. All other observed odd-parity levels belong to the
configuration $3d^6 4p$. Note that the ground state belongs to the
configuration $3d^6 4s$. It means that corresponding transitions are
of $4s \rightarrow 4p$ type, while transition to the level
$^6P^o_{7/2}$ is of $3d \rightarrow 4p$ type. The relativistic
corrections increase binding energy of $s$-electrons and decrease
binding energy of $d$-electrons. Because of that the $q$-factor for
$^6P^o_{7/2}$ level is of the opposite sign to the $q$-factors of
all other observed levels. The existence of the $q$-factors of both
signs makes Fe~\textsc{ii} ion so attractive for the search of
$\alpha$-variation.

Initially there was some controversy around this level. In Moore's
tables \cite{Moo58} this level was erroneously marked as
$3d^6({}^7\!S)4p\,{}^6\!P^o_{7/2}$. It was an obvious misprint
because there is no term $^7\!S$ in configuration $3d^6$, but there
is such term in configuration $3d^5 4s$. Still, this assignment was
used in the first calculation of Fe~\textsc{ii} \cite{MWF01a} for
the identification of this level and a positive $q$-factor was
obtained. In the next calculation \cite{DFK02} this error was
noticed and negative $q$-factor for this level was reported.

One of the main arguments which allowed to assign this level to the
configuration $3d^5 4s4p$ was based on the analysis of $g$-factors.
The experimental $g$-factor for this level is 1.68
\cite{Moo58,NIST}, while all levels of the configuration $3d^6 4p$
have significantly smaller $g$-factors. On the other hand, the Lande
value of the $g$-factor for the level $^6P_{7/2}$ for the pure
$LS$-coupling scheme is 1.714. Calculated value of the $g$-factor
appeared to be very close to the Lande value (see \tref{tab_w_g_q}).
This was used in Ref.\ \cite{DFK02} to estimate configuration mixing
for this level to be about few percent and calculate corresponding
correction to the $q$-factor:
 \begin{align}\label{dq_g}
 \delta q_g\approx +180.
 \end{align}

Since the publication of the paper \cite{DFK02} the new NIST tables
for Fe~\textsc{ii} became available \cite{NIST}. These tables
confirm assignment of the level $^6P^o_{7/2}$ at 62172 cm$^{-1}$ to
configuration $3d^5 4s4p$. Moreover, the mixing of the
configurations $3d^5 4s4p$ and $3d^6 4p$ for this level at a few
percent was also independently confirmed from the analysis of the
branching ratios \cite{PDN02}. Therefore, we conclude that
application of the correction \eqref{dq_g} is justified. The
correction of the opposite sign has to be applied to the closest
interacting level $^4F^o_{7/2}$.

At present it is impossible to reproduce this configurational mixing
in the \textit{ab initio} calculations because it is very sensitive
to the relative position of the levels. The non-diagonal matrix
elements which cause this mixing are very small, on the order of 10
cm$^{-1}$, and mixing takes place only when the splitting between
interacting levels is about 100 cm$^{-1}$ or less. However, the
accuracy of the present theory is insufficient to reproduce the
energy intervals between the levels of different configurations with
such accuracy (see \tref{tab_w_g_q}). Because of that we still have
to introduce correction \eqref{dq_g} semi-empirically. If more
accurate experimental data on $g$-factors and the branching ratios
become available they may help to refine this analysis.

\section{conclusions}\label{conclusions}

Our final results for the $q$-factors of the astrophysically
important levels of Fe~\textsc{ii} ion are given in
\tref{tab_final}. In addition to the large-scale 7-electron CI
calculation for Dirac-Coulomb ``no pair'' Hamiltonian, we calculated
corrections due to the Breit interaction and core-valence
correlations. Both corrections appeared to be larger, than was
anticipated in previous calculation \cite{DFK02}. Nevertheless our
calculations confirm previous conclusion that the $q$-factors of the
levels of the configuration $3d^6 4p$ are large and positive, while
the state $^6\!P^o_{7/2}$ at 62172 cm$^{-1}$ belongs to the
configuration $3d^5 4s4p$ and has large negative $q$-factor.

\begin{table}[htb]
\caption{Recommended values for $q$-factors in respect to the ground
state, $q_\mathrm{rec}=q_\mathrm{val}+\delta q_{br} +\delta
q_\mathrm{cv} +\delta q_g$, with uncertainties in parentheses (in
cm$^{-1}$). Valence contribution $q_\mathrm{val}$ corresponds to the
longest basis set $[7spdf]$ from \tref{tab_w_g_q}. Breit correction
$\delta q_{br}$ and core-valence correlation correction $\delta
q_\mathrm{cv}$ are taken from Tables~\ref{tab_breit}
and~\ref{tab_core}. Finally, $g$-factor correction $\delta q_g$ is
given by \Eref{dq_g}. }

\label{tab_final}

\begin{ruledtabular}
\begin{tabular}{llrrrrrr}
&&&&&&\multicolumn{2}{c}{$q_\mathrm{rec}$}\\
Conf.& Level
 &\multicolumn{1}{c}{$q_\mathrm{val}$}
 &\multicolumn{1}{c}{$\delta q_{br}$}
 &\multicolumn{1}{c}{$\delta q_\mathrm{cv}$}
 &\multicolumn{1}{c}{$\delta q_g$}
 &\multicolumn{1}{c}{This work}
 &\multicolumn{1}{c}{Ref.\cite{DFK02}}
\\
\hline
$3d^6 4p$   & $^6\!D^o_{9/2}$   &$ 1334    $&$ 19$&$ 59$&$    0 $&$ 1410\,(60)\,\,$&$ 1330\,(150)  $\\
$3d^6 4p$   & $^6\!D^o_{7/2}$   &$ 1488    $&$ 11$&$ 36$&$    0 $&$ 1540\,(40)\,\,$&$ 1490\,(150)  $\\
$3d^6 4p$   & $^6\!F^o_{11/2}\!$&$ 1479    $&$ 11$&$ 55$&$    0 $&$ 1550\,(60)\,\,$&$ 1460\,(150)  $\\
$3d^6 4p$   & $^6\!F^o_{9/2}$   &$ 1607    $&$ -3$&$ 55$&$    0 $&$ 1660\,(60)\,\,$&$ 1590\,(150)  $\\
$3d^6 4p$   & $^6\!P^o_{7/2}$   &$ 1134    $&$ 13$&$394$&$    0 $&$ 1540\,(400)   $&$ 1210\,(150)  $\\
$3d^6 4p$   & $^4\!F^o_{7/2}$   &$ 1240    $&$ 47$&$449$&$\!-180$&$ 1560\,(500)   $&$ 1100\,(300)  $\\
$3d^5 4s4p$ & $^6\!P^o_{7/2}$   &$\!\!-1506$&$131$&$168$&$   180$&$\!-1030\,(300) $&$\!-1300\,(300)$\\
\end{tabular}
\end{ruledtabular}
\end{table}

We found out that the Breit interaction between valence electrons is
of the same order of magnitude as the core-valence Breit
interaction. Methods accounting for MBPT corrections developed so
far do not allow us to treat core-valence correlations accurately
for such complicated systems as Fe~\textsc{ii}. Besides, we were
unable to reproduce in the frame of the CI method strong mixing
between the level $^6\!P^o_{7/2}$ of configuration $3d^5 4s4p$ and
nearly located levels of configuration $3d^6 4p$. Corrections to
$q$-factors due to this mixing were found semi-empirically using
$g$-factor analysis \cite{DFK02}. This analysis is qualitatively
confirmed by the analysis of the branching ratios \cite{PDN02}.
Considering the two problems mentioned above as a main source of
uncertainties we can roughly estimate the final uncertainties,
$\Delta$, of the $q$-factors presented in~\tref{tab_final} as
$$\Delta \simeq \sqrt{ (\delta q_\mathrm{cv})^2 + (\delta q_g)^2}.$$
As is seen from~\tref{tab_final} the uncertainties for the
$q$-factors for the low-lying astrophysically important states are
at the level of 4\% while for higher-lying levels they attain 30\%.
Further development of the CI+MBPT method allowing to treat
core-valence correlations more accurately and appearance of more
precise experimental data on $g$-factors and transition rates for
these levels would be very helpful in improving the accuracy of the
$q$-factors.

\section{acknowledgments}

This work was supported in part by the Russian Foundation for Basic
Research under Grant Nos. 05-02-16914-a, 06-02-16489, and
07-02-00210-a; by the Federal Agency for Science and Innovations
under Grant NSh~9879.2006.2; by St.~Petersburg State Scientific
Center, and by DFG Grants SFB 676 Teilprojekt C and RE 353/48-1. SP
and KK thank Hamburg University for hospitality.


\end{document}